\documentclass{JHEP3}
\preprint{\\KA-TP-09-06\\MCNET-09-13}

\usepackage[utf8]{inputenc}

\usepackage{cite}
\usepackage{epsfig}
\usepackage{color}
\usepackage{rotating}
\usepackage{lscape}

\usepackage{graphics}
\usepackage{inputenc}
\usepackage{xspace}
 \renewcommand\email[1]{{\scriptsize\tt\href{mailto:#1}{#1}}}

\newcommand{\M}{\ensuremath{\mathbf{M}}}
\newcommand{\G}{\ensuremath{\mathbf{\Gamma}}}
\newcommand{\Sv}{\ensuremath{\mathbf{S}}}
\newcommand{\Ng}{\ensuremath{N_g}}
\newcommand{\Nc}{\ensuremath{N_c}}
\newcommand{\Nq}{\ensuremath{N_q}}
\newcommand{\Nqbar}{\ensuremath{N_{\overline{q}}}}
\newcommand{\Nparton}{\ensuremath{N_{\mbox{\small{p}}}}}

\newcommand{\q}{\ensuremath{q}}
\newcommand{\qbar}{\ensuremath{\overline{q}}}

\newcommand{\Nbasis}{N_{\mbox{\small{basis}}}}
\newcommand{\Tr}{\mbox{Tr}}

\newcommand{\as}{\ensuremath{{\alpha}_{s}}}

\def\beq{\begin{equation}}
\def\eeq{\end{equation}}
\def\beqa{\begin{eqnarray}}
\def\eeqa{\end{eqnarray}}

\newcommand{\eqref}[1]{Eq.~(\ref{#1})\xspace}

\skip\footins = 1\bigskipamount plus 2pt minus 4pt                              
\title{\boldmath Color structure for soft gluon resummation - a general
recipe }

\author{Malin Sjödahl\\
Institut für Theoretische Physik, Universität Karlsruhe,
Wolfgang-Gaede-Str. 1, Physikhochhaus, 76131 Karlsruhe, Germany\\
  E-mail: \email{malin@particle.uni-karlsruhe.de}
}
  
  \abstract{
A strategy for calculating the color structure needed for soft gluon 
resummation for processes with any number of colored partons is 
introduced using a $N_c \rightarrow \infty$ inspired basis.
In this basis a general formalism can be found at the same time as the
calculations are simplified.

The advantages are illustrated by recalculating the soft anomalous
dimension matrix for the processes $gg \rightarrow gg$, $q\qbar
\rightarrow q \qbar g$ and $q\qbar \rightarrow ggg$.
}

\begin{document}
 
 
\section{Introduction}
\label{sec:intro}

The strong force comes with the problem of
being precisely strong. 
Although the coupling constant of QCD is small enough
for perturbation theory to make sense at all, it is large enough to cry
out for higher order corrections for many processes, and in some regions
of phase space, large enough to invalidate a fixed order calculation. 

This is the case in the collinear region, where a large logarithm
compensates for the moderate smallness of $\as$,
and similarly in the soft region where there is a large effective phase space
$\sim \log (\mbox{hard scale}/\mbox{soft resolution scale})$ in
transverse momentum. In these regions resummation methods are needed. 
In the collinear DGLAP region \cite{Lipatov:1974qm,Gribov:1972ri,Altarelli:1977zs,Dokshitzer:1977sg}, where the emission can be seen as coming from
one parton, the color structure is trivial and Sudakov form factors can
be used to describe no-emission \textit{probabilities}. 

Unfortunately, the
strong force is not only strong, it is also complicated, in the sense of
being non-Abelian. In the soft region, where emissions have
contributions from branchings of different partons this complicates
matters.
The real emission coming from the interference term of emission
off parton $i$ and emission off parton $j$, is canceled by the virtual gluon
exchange between parton $i$ and $j$. (Using Feynman gauge 
self-energy type diagrams can be neglected, i.e. $i \ne j$.)

Under the assumption that emissions strongly ordered in transverse
momentum dominate, all leading logarithms in 
(hard scale/soft scale) from virtual corrections
exponentiate and can be resummed. However, since these gluon exchanges affect
the color structure, the exponentiation must be done at the amplitude
level. Thus a no-emission \textit{amplitude}
\begin{equation}
  \label{eq:M}
  \M=\exp\left(  -\frac{2}{\pi}
	 {\displaystyle\int\limits_{Q_{0}}^{Q}}
	 \alpha_{s}({k'}_{\perp}) \frac{d{k'}_{\perp}}{{k'}_{\perp}} \G \right)  \M_{0},
\end{equation}
can be derived. In the above, $\M_{0}$ is the undressed hard scattering
amplitude as a vector in color space and $\G$ is a matrix in color
space, describing the effect of exchanging gluons between the various partons,

\begin{equation}
  \G=\sum_{i < j}\Omega_{ij}{\mathbf{C}^{ij}}.
  \label{eq:G}
\end{equation}
Here $\mathbf{C}^{ij}$ describes the color algebra part and
$\Omega_{ij}$ 
contain the azimuth and rapidity momentum integral over the exchanged
gluon $k'$,

\begin{equation}
  \Omega_{ij}=
  -\frac{1}{2} 
  (-1)^l 
  \left[
    \int_{\Omega} \frac{dy' d \phi'}{2 \pi} 
    \frac{{k'}_{\perp}^2 p_i \cdot p_j}{2 p_i \cdot  k' k'\cdot p_j}
    -\frac{1}{2}(1-s_{ij})i \pi
    \right]
  \label{eq:PhaseSpace}
\end{equation}
with $s_{ij}=-1$ if the partons $ij$ are both incoming or both outgoing, 
and $1$ otherwise, and  
$l$ counts how many of the involved partons
which are quarks in the initial state, anti-quarks in the final state or
gluons, assuming the convention in \eqref{eq:gggSign} for the triple gluon
vertex.
In the above equation the $i \pi$-terms, coming from Coulomb gluon exchange, 
would give rise to an unobservable phase in an Abelian theory. 
For a non-Abelian theory they do, however, enter in a physically relevant way.

In general the color basis used need neither be orthogonal or
normalized. In fact, it will be seen below that the calculations simplify
significantly in a special basis which is not. For a non-orthonormal
basis, the matrix of scalar products $\Sv$, calculated by summing over
quark, anti-quark and gluon indices $a,b,c,...$ 
\begin{equation}
  \Sv_{mn}=<C^m,C^n>=\sum_{a,b,c...}C^m_{abc...}(C^n)^{*}_{abc...},
  \label{eq:SP}
\end{equation}
is needed. (Note that $C^m$ above is a basis tensor in color space,
whereas $\mathbf{C}^{ij}$ in \eqref{eq:G} are matrices in this basis,
describing the effect of gluon exchange between parton $i$ and $j$.) 
The physical no-emission probability is given by 
$\sigma= \M^\dagger \Sv \M$. As an aside it is pointed out that
scalar products between tensors corresponding to linear combinations of
color structures of Feynman diagrams with real coefficients are real.

In the simple case of
$q_1 q_2\rightarrow q_3 q_4$ the vector space containing the color structure has
only two dimensions, and the basis vectors are often taken to be the
``$t$-channel singlet octet basis'',

\begin{eqnarray}
  C^1_{q_1 q_2 q_3 q_4}&=& \delta_{q_1 q_4}  \delta_{q_2 q_3} \nonumber  \\
  C^2_{q_1 q_2 q_3 q_4}&=& 
  t^g_{q_4 q_1} t^g_{q_3 q_2}=
  \frac{1}{2}\left[ \delta_{q_1 q_3}  \delta_{q_2 q_4} \
    -\frac{1}{\Nc} \delta_{q_1 q_4}  \delta_{q_2 q_3}\right].
\end{eqnarray}
In this case the issue of keeping track of the color structure
amounts to a moderate complication. However, already for $gg
\rightarrow gg$ a six dimensional vector space is needed 
(reducing to a five
dimensional space for $\Nc=3$) and for $gg \rightarrow ggg$ there are 22
different color states to keep track of (reducing to 16 for $\Nc=3$)
\cite{Kidonakis:1998nf, Sjodahl:2008fz}. In
the later case, to keep track of the change in color structure as a
result of virtual gluon exchange between a pair of partons, one 
naively - without using further symmetries,
thus needs
to calculate the effect of gluon exchange on 22 different color
states, and then decompose the result into the 22 different color tensors by
taking scalar products, implying in total $22^2$ scalar product. 
(This number may be reduced, for example by
using the fact that the soft anomalous dimension matrices are symmetric
if stated in orthonormal bases \cite{Seymour:2005ze, Seymour:2008xr}.)
The color structure thus gives rise to a major computational
complication, and so far the soft anomalous dimension matrices have only
been calculated for the $2 \rightarrow 2$ processes 
\cite{Botts:1989kf,Sotiropoulos:1993rd,Contopanagos:1996nh,Kidonakis:1998nf,Oderda:1999kr,Appleby:2003hp,Dokshitzer:2005ig} and the
$2\rightarrow 3$ processes \cite{Sjodahl:2008fz,Kyrieleis:2005dt}.
(For observable related and experimental work, see for example 
\cite{Oderda:1998en,Berger:2001ns,Berger:2002ig,Appleby:2002ke,Derrick:1995pb,Abe:1997ie,Abe:1998ip,Abbott:1998jb,Adloff:2002em}.)

If one is only interested in a fixed order expansion, for example as in
\cite{Forshaw:2008cq}, there is no need to choose an explicit basis. 
Indeed the soft anomalous dimension matrix can be written down in a
compact basis-independent way for any number of partons, both at one-lopp and
two-loop order \cite{ MertAybat:2006mz}. 
Similarly,
for the purpose of deriving general theoretical properties it is often wiser to
stay basis independent, and several interesting results have recently
been derived without explicit basis choices \cite{Gardi:2009qi,
  Becher:2009cu, Becher:2009qa,Becher:2009kw, Mitov:2009sv,Seymour:2008xr}. 
However, to actually perform the numerical exponentiation of
\eqref{eq:M}, to obtain all-order results, an explicit basis is needed.

It is thus clearly desirable to find a simplifying general strategy.
Especially, a unified formalism is needed for the long term goal to
incorporate non-leading color effects in event generators. The major
current event generators all work in the leading $\Nc$ limit
\cite{Sjostrand:2006za,Sjostrand:2007gs,Corcella:2002jc,Bahr:2008pv,Gleisberg:2003xi}. 
This means
that the color structure is decomposed into leading $\Nc$ contributions,
using \eqref{eq:tata} and \eqref{eq:fd} below. Color suppressed interference
terms between different color structures are neglected. It was argued a
long time ago that for gluon amplitudes with fixed power of $\as$ these
terms are suppressed by $1/\Nc^2$ 
\cite{'tHooft:1973jz}. 
However,
there may in general be many suppressed
terms. As an example consider $\Ng-2$ gluons attached in a row to one
gluon line, giving in total $\Ng$ gluons. (For $\Ng$ up to five, all tree
level graphs have this topology.) The squared amplitude is given by
\begin{equation}
\Nc^{\Ng-2}(\Nc^2-1).
\label{eq:all2}
\end{equation} 
If the diagram is decomposed into different color topologies (which are
orthogonal in the $\Nc \rightarrow \infty$ limit) the sum of the parts
squared separately is 
\begin{equation}
  \frac{1}{\Nc^{\Ng}}\left[(\Nc^2 -1)^{\Ng} +(-1)^{\Ng}(\Nc^2-1) \right].
\label{eq:parts2}
\end{equation} 
When $\Nc \rightarrow \infty$ both expressions grows as $\Nc^{\Ng}$
and their ratio approaches one. However for finite $\Nc$ the difference grows
with $\Ng$ and 
already for $\Ng=4$, if $\Nc=3$, \eqref{eq:parts2} is only 19/27 of 
\eqref{eq:all2}, \cite{Gustafson:1982ws}. For 7 gluons \eqref{eq:parts2} 
is less than 50\% of \eqref{eq:all2}.

The method suggested in this paper for dealing with the color structure
of multi-parton
processes is developed with the resummation of soft gluons in
mind, but clearly, as it describes the effect of gluon exchange on any
colored amplitude, it may also prove useful for
NLO (and higher order) corrections to amplitudes with (many) colored partons.

The results may also be used to calculate effects
stemming from the non-global nature of most observables, the ordinary
``non-global logs'' \cite{Dasgupta:2001sh,Dasgupta:2002bw}, as well as the
color suppressed
``super leading logarithms'' carrying extra powers of $
\log(\mbox{hard scale/soft scale})$, suggested to enter at order $\as^4$ 
in perturbation theory
\cite{Forshaw:2006fk,Forshaw:2008cq,Forshaw:2009sf}.
Indeed, as the non-global logarithms originate from real radiation
outside an experimental exclusion region, to calculate the contribution
from $n$ emissions outside the exclusion region requires the soft
anomalous dimension matrices for processes containing $n$ additional partons.

As the two-loop soft anomalous dimension matrices have been proven to be
proportional to the one-loop results (for processes with any number of
colored and uncolored massless external legs), 
the present method can trivially be used also for two-loop 
anomalous dimension matrices \cite{ MertAybat:2006mz}.
Recently it has been suggested that similar results also hold
for the three-loop anomalous dimension matrices and that they may hold to
any order, as long as the partons remain massless 
\cite{Gardi:2009qi, Becher:2009cu, Becher:2009qa}. 
For massive external legs this simple relation breaks down 
\cite{Becher:2009kw, Mitov:2009sv}. 

The layout of this paper is as follows: First the formalism for
constructing a basis is described in section \ref{sec:Construction},
and computational rules for gluon exchange in this basis are
derived in section \ref{sec:Rules}. To illustrate the advantages with
the constructed bases, the soft anomalous dimension matrices for 
$gg \rightarrow gg$, $q\qbar \rightarrow q\qbar g$ and 
$q \qbar \rightarrow ggg$ are recalculated in section \ref{sec:Example}. 
Finally some concluding remarks are made in section \ref{sec:Conclusions}.

\section{General basis formalism}
\label{sec:Construction}

\subsection{Construction of a general basis}

Previous strategies for dealing with the color structure needed for
resummation of soft gluons have lately been based on multiplet
decomposition for finding a basis
\cite{Dokshitzer:2005ig,Sjodahl:2008fz}. 
In this way symmetry properties are 
exploited to construct complete orthogonal bases 
(which easily can be normalized).
Clearly, using an orthogonal basis has advantages. The result is easy
to interpret and the matrix of scalar products between basis vectors is
diagonal.

A disadvantage of the multiplet strategy is, however, that increasingly 
complicated projection operators need to be used, 
and no closed form for deriving these projection operators exists 
(to the knowledge of the author).

Another complication is that the projection operators, which tend to be
expressed in terms of the symmetric and anti-symmetric structure
constants $f_{abc}$ and $d_{abc}$, need increasingly complicated
computational rules for contraction of indices, that is,
computational rules involving more and more $f$'s and
$d$'s. Alternatively, the structure constants can be reexpressed in
terms of the generators of the fundamental representation $t^g_{q_1q_2}$,

\begin{equation}
  \left.
      \begin{array}{l}
        i f_{abc} \\
        d_{abc}  
      \end{array} \right\}
  =2(\Tr[t^a t^b t^c] \mp \Tr[t^b t^a t^c])=
  2(t^a_{q_1 q_2} t^b_{q_2 q_3}t^c_{q_3 q_1} \mp 
  t^b_{q_1 q_2} t^a_{q_2 q_3}t^c_{q_3 q_1}).
  \label{eq:fd}
\end{equation}
In this case any scalar product, of arbitrarely complicated color
tensors, can be calculated using the gluon index contraction relation

\begin{equation}
t^g_{ca} t^g_{db}
=\frac{1}{2}(\delta_{ad}\delta_{bc}-\frac{1}{N_c}\delta_{ac}\delta_{bd}).
\label{eq:tata}
\end{equation}
However, the expression for the color structure tensor 
will contain $2^{(\mbox{\small{\# of $f$'s and $d$'s}})}$ terms, and the scalar 
product of the tensor with itself thus $2^{2(\mbox{\small{\# of $f$'s and $d$'s}})}$ 
terms which each has to be contracted separately.

An alternative strategy would be to construct a basis by starting from a
sufficient number of arbitrarily chosen color tensors, 
or by exploiting possible symmetries. 
This will work well for a small vector space, cf.
\cite{Kyrieleis:2005dt}, but will tend to give very lengthy expressions
for the basis vectors if Gram-Schmidt orthogonalization is used
for a large vector space. 
On the other hand, if the basis vectors are not made orthogonal the
decomposition of color structures resulting after gluon exchange will
in general be cumbersome. 
(This complication is circumvented in the special non-orthogonal basis
suggested below).
In addition it has to be proved that the basis
actually span the relevant space.

These issues make it worth exploring other strategies for constructing
the basis in the general case of any number of colored and uncolored 
partons.
The basis clearly has to span the relevant space. It may seem desirable
to find an orthogonal (normalized) basis, but it will be seen below that
using a special non-orthogonal, non-normalized basis significantly
diminishes
the computational effort, mainly since the state obtained after gluon
exchange is immediately, i.e. without taking scalar products, a linear
combination of basis states. There is thus no need
for calculating ${\Nbasis}^2$ scalar products for every possible 
gluon exchange. 

The solution is to use a basis inspired by the $\Nc \rightarrow \infty$ limit. 
In the case of infinitely many colors, two color lines in a Feynman
diagram are never the same, and gluons may be represented by
two color lines going in opposite directions. In this case, all possible
color structures can  be represented by all ways of connecting
incoming and outgoing color lines. The strategy suggested here is thus
similar to methods used in
\cite{Paton:1969je,Mangano:1987xk,Berends:1987cv,Nagy:2007ty}. 
Especially it is noted that the bases suggested here for resummation
are similar to the color structure treatment suggested in
\cite{Nagy:2007ty} to deal with real parton emission in event generators.
For $\Nc=\infty$ 
the scalar product between different color topologies, divided by the 
scalar product of a topology with itself, equals zero. 
However, for finite $\Nc$ there are scalar product terms which are
suppressed only by $1/\Nc$.

Another important property of the bases constructed in the
aforementioned way is that they
are completely democratic w.r.t. different quarks, different anti-quarks
and different gluons. This implies, for example, that once the
effect of gluon exchange between the gluons $g_1$ and $g_2$ has been
calculated, the effect of gluon exchange between any other gluons can be
obtained by relabeling of indices i.e. renumbering of basis tensors. One
therefore never needs to calculate more than six different exchanges $gg$,
$qq$, $\qbar \,\qbar$, $gq$, $g\qbar$ and $q \qbar$. In addition, it
will be seen below that the color structure after a gluon exchange on a
given color topology is a 
linear combination of at most four different basis tensors. 
The soft anomalous dimension matrices will thus be relatively sparse in
the suggested bases, which should simplify numerical exponentiation.

It is also worth stressing that the suggested bases are well suited for
comparison to the $\Nc \rightarrow \infty$ limit, as the bases are easy
to interpret and the soft anomalous
dimension matrices will turn out to be diagonal in this limit.
This implies that they are
ideal for comparison to the radiation pattern obtained from event
generators tending to work in the 
$\Nc \rightarrow \infty$ limit
\cite{Sjostrand:2006za,Sjostrand:2007gs,Corcella:2002jc,Bahr:2008pv,Gleisberg:2003xi}.

The reduction in calculational effort for the soft anomalous dimension
matrix with the suggested basis is thus
threefold. There is \textit{no} need to calculate scalar products,
reducing the computational effort with a factor 
$\sim \Nbasis^2$ from the number of scalar products
\textit{and} a factor 
$2^{(\mbox{\small{\# of $f$'s and $d$'s}})}$
from the number of terms in each of the 
scalar products, assuming \eqref{eq:fd} is used.
Furthermore there are at most six, as compared to 
$\Nparton(\Nparton-1)/2$ for $\Nparton$ external particles, 
different gluon exchanges to keep track of, the
others are related by relabeling of indices.

Unfortunately this does not quite remove the bad scaling of the problem
with the number of partons, as instead of having to calculate
$\sim \Nbasis^2$ scalar products for each contribution to the soft
anomalous dimension matrix, one
has to calculate $\sim \Nbasis^2$ scalar products between the basis vectors,
as they are only orthogonal in the $\Nc \rightarrow$ infinity limit. 
However, this only has to be done once. In addition calculating 
scalar products using \eqref{eq:fd} and \eqref{eq:tata} gives just one, 
as opposed to $2^{(\mbox{\small{\# of $f$'s and $d$'s}})}$, different terms.

What remains is thus a scaling of type $\Nbasis^2$. Very roughly
speaking $\Nbasis$ tends to grow as $\Nparton!$, cf. section 
\ref{sec:QuarksOnly}-\ref{sec:Both}.
But, bearing in mind that only the topology of the color contraction,
and not the labeling of indices is important for the scalar product,
should naively reduce the $(\Nparton!)^2$ scaling by a factor 
$\sim \Ng!\Nq! \Nqbar!$ from the
number of ways of labeling the indices. What remains is then a factorial
growth for processes with only gluons.

Note however, that for processes with many enough external partons, the major
computational effort will not lie in finding an expression for the soft
anomalous dimension matrix, but in numerical exponentiation of the
obtained result. As numerical matrix exponentiation scales with the cube of
the matrix size, and the number of basis vectors tends to grow
factorially with the number of partons, calculations with more than ten
particles seem unlikely. For practical implementations, it is also worth
pointing out that the number of basis vectors highly depend on the kinds 
of partons involved. For processes with no external gluons and
$\Nq=\Nqbar=\Nparton/2$ partons, the number of basis vectors is 
$(\Nparton/2)!$ whereas for processes with only external gluons the size
of the basis tens to grow rather as $\Nparton!/e$, cf. section
\ref{sec:QuarksOnly} and \ref{sec:Both}.

That a basis constructed in the above described way is complete for 
$\Nc=\infty$ is clear
from the fact that it represents all possible color topologies. For
finite $\Nc$, some of the color tensors may be linearly dependent, and the
basis over-complete, but it will still span the space. One way of
thinking of the reduction in dimension of the color space is to note
that tensors corresponding to multiplets which are anti-symmetric in
more than $\Nc$ quark indices are not possible. Requiring
that a color decomposition should be valid for all $\Nc$ defines a unique
decomposition of a $\Nc=3$ tensor.

Another way of convincing oneself that the above bases are complete, is to 
note that \textit{every} internal gluon line in any Feynman diagram can be
removed by first using \eqref{eq:fd} to remove the triple gluon vertices
and then \eqref{eq:tata} to remove gluon propagators. In this way any
Feynman diagram, tree level or not, can be decomposed into color
structures containing no gluon propagators. What remains is a
linear combination of color structures containing internal quark lines,
external quarks, external anti-quarks and external gluons. 
That is, a linear combination of terms of precisely the form obtained by
first splitting all gluons to $\q \qbar$-paris, and then
connecting quark and anti-quark lines in all possible ways.

Below, the construction of basis tensors will be investigated in more
detail, first in the special case of external quarks only, then for
external gluons only, and finally in the general case of both.

Before moving on we note that from the color algebra point of view there 
is no difference between an outgoing quark and an incoming anti-quark,
from here on simply collectively referred to as quark, 
or an incoming quark and outgoing anti-quark, from now on referred to as 
anti-quark. Opposite conventions may be used elsewhere. In addition
the placing of quark and anti-quark indices on the fundamental
generators may be varied.

\subsection{The quarks only case}
\label{sec:QuarksOnly}

Finding a basis in the case of only external quarks is trivial. The basis just
consists of all possible ways of connecting quarks and anti-quarks. 
For $\Nq=\Nqbar$ quarks (clearly, for each incoming quark line there is also an
outgoing) this can be done in 

\begin{equation}
  \Nbasis=\Nq!
\end{equation}
ways.
The squared norm of these basis vectors, calculated using \eqref{eq:SP},
is equal to $\Nc^{\Nq}$.

To denote the tensors the notation 

\begin{equation}
  ({\qbar}_1 q_3)({\qbar}_2 q_4)=\delta_{q_1 q_3}\delta_{q_2 q_4}
\end{equation}
is used.
A complete basis for $q_1 q_2 \rightarrow q_3q_4$ is thus the tensors
$({\qbar}_1 q_3)({\qbar}_2 q_4)$ and $({\qbar}_1 q_4)({\qbar}_2 q_3)$.
In fact this is the basis used in \cite{Sotiropoulos:1993rd}.

\subsection{The gluons only case}
\label{sec:GluonsOnly}

To construct the basis in the case of gluons only, closed quark loops with 
external gluons attached are used. 
For example, for four gluons, all gluons may be connected to the
same quark line giving $(4-1)!=6$ topologically different
diagrams. Alternatively the gluons may be connected two and two in three
different ways. Indeed the color space also has nine dimensions, 
however, only half of the linear combinations of the six
fully connected topologies are physical, due to the fact that
quarks and anti-quarks enter QCD on equal
footing. Therefore, if, in a quark loop, a quark is going around in one
direction, the topology with the quark going around in the opposite
direction (i.e. the gluon index order is reversed) must also contribute. 

More explicitly, introducing the notation
\begin{equation}
  (g_1 g_2...g_{\Ng})
  =\Tr[t^{g_1} t^{g_2} ... t^{g_{\Ng}}]
  =t^{g_1}_{q_1 q_2}t^{g_2}_{q_2 q_3}...t^{g_{\Ng}}_{q_{\Ng} q_1},
\end{equation}
to denote $\Ng$ gluons attached clockwise in the order $g_1...g_{\Ng}$ on 
a quark line, we note that the physical linear combinations must be
\begin{equation}
  (g_1 g_2...g_{\Ng})+(-1)^{\Ng}(g_{\Ng}...g_2 g_1).
  \label{eq:PhysG}
\end{equation}
To understand the sign, decompose any tree level Feynman diagram with
only gluons using \eqref{eq:fd} and \eqref{eq:tata}. The
result is a sum of color structures where the $\Ng$ gluons are attached
in different orders to the quark-line. For a specific order, the
anti-cyclic order is obtained by reversing the direction of the
quark-line in \textit{every} vertex, i.e. taking the other term in
\eqref{eq:fd} everywhere. This gives a factor $(-1)^{\Ng-2}$ as there
are $\Ng-2$ vertices, explaining the sign
in \eqref{eq:PhysG}.

Thus, in the case of $gg \rightarrow gg$, only six color tensors are needed
(for general $\Nc$). This explains the observation that some
tensors decouple for $gg \rightarrow gg$ and $gg \rightarrow ggg$ 
\cite{Kidonakis:1998nf,Oderda:1999kr,Sjodahl:2008fz}.

The problem of constructing the $\Ng$-gluon basis in the general case thus
boils down to:

\begin{list}{}{\setlength\leftmargin{10mm}}
\item[(1)]{
Find all the ways of grouping the $\Ng$ gluons such that each group
contains at least two gluons. (Groups with only one gluon would correspond
to the color structure $t^g_{qq}=0$.) For four external gluons the possible
groupings are thus $\{4\}$ and $\{2,2\}$.}

\item[(2)]{ For each fully connected grouping, such as $\{4\}$, 
find all physical different ways of arranging the gluons. 
For $\Ng$ gluons this gives
$(\Ng-1)!/2$ different color tensors where
the factor $1/2$ is present since only one combination of the cyclic 
and anti-cyclic ring is physical.
}

\item[(3)]{ For disconnected groupings, such as $\{2,2\}$,} 
\end{list}
\begin{list}{}{\setlength\leftmargin{20mm}}
\item[(3a)]{
 Find separately, for each
subgroup, all physically different ways of arranging the gluons.
}
\item[(3b)]{
Distribute the gluon indices $\{g_1....g_{\Ng}\}$ in all possible
ways among the different subgroups.
}
\item[(3c)]{
 Combine the different sub-groupings in all possible ways, taking
into account that, if all gluon indices are equal, two groupings do
actually correspond to the same physical state. For example the
subgrouping $\{\{g_1,g_2\},\{g_3,g_4\}\}$ and $\{\{g_3,g_4\},\{g_1,g_2\}\}$ are equal.
}
\end{list}
Following this recipe a complete basis describing the color structure
for any number of external gluons can be constructed. 

Neglecting the issue of physical linear combinations, the possible color
tensors coincide with the color tensors obtained by replacing each gluon
with one quark and one anti-quark line,
with the important exception that contractions between a $\q\qbar$ pair
corresponding to the same gluon are disallowed. The problem of finding all such
topologies is equivalent to the number
of ways of mapping $N$ elements to each other without mapping a single one to
itself, which has a known solution 

\begin{equation}
N! \sum_{i=0}^N \frac{(-1)^i}{i!} \rightarrow \frac{N!}{e}.
\label{eq:NoSelf}
\end{equation}
The convergence to $N!/e$ is very quick, rounding off to the closest
integer works already for $N=1$.

Note however, that this just gives the total number of linearly
independent color tensors (for $\Nc=\infty$). As mentioned above, only
tensor combinations where quarks and anti-quarks enter on equal footing
are physical. 
For every quark ring participating in building up a color tensor, the
corresponding anti-quark ring has to be added. This reduces
the number of physical tensors of a certain topology, such as $\{3,2\}$
with a factor $(1/2)^{\mbox{\small{\# rings building up the
      tensor}}}$, that is $(1/2)^2$ for $\{3,2\}$.

As the number of fully connected color topologies, where all $\Ng$ gluons are
attached to the same quark-line equals
$(\Ng-1)!$, the fraction of color tensors corresponding to fully connected
diagrams is roughly 
$e/\Ng$,
again ignoring the issue of physical tensor combinations.
Tree level QCD Feynman diagrams with only
external gluons (more generally, no gluon propagator between quarks)
always correspond to linear combinations of fully connected diagrams,
(i.e. diagrams where all gluons are connected to the same quark-line)
and are the only diagrams presently included in major event generators. When
considering only physical topologies the ratio of fully connected to
disconnected graphs changes slightly to the advantage of the fully
connected graph, as the factor 
$(1/2)^{\mbox{{\small{\# rings building up the tensor}}}}$ 
hits the disconnected topologies harder.

The norm of the color tensor with all gluons attached to a quark going
around in \textit{one} direction is given by

\begin{equation}
\left(\frac{1}{2\Nc}\right)^{\Ng}
\left[(\Nc^2-1)^{\Ng}+(-1)^{\Ng} (N_c^2-1) \right].
\label{eq:RingNorm}
\end{equation}
The physical tensors, being sums of gluons attached to
rings with quarks going around in opposite directions, contain mixed terms
as well, these are however relatively suppressed, and for large $\Ng$ or
large $\Nc$ \eqref{eq:RingNorm} is a good approximation. 
Note that the
norm grows as $\Nc^{\Ng}$, which is to be expected considering the
$\Nc=\infty$ limit. 
It turns out, however, that it is easier to stick to the non-normalized
versions of the color tensors.

\subsection{The case of both quarks and gluons}
\label{sec:Both}

In the general case of both external quarks and gluons the basis may be
constructed by:

\begin{list}{}{\setlength\leftmargin{10mm}}
\item[(1)]{Connect the quark lines to each other in all possible ways, giving
$\Nq!$ possibilities.
}
\item[(2)]{For $i=1,2,... \Ng-2,\Ng$ attach $i$ of the $\Ng$ gluons to the 
quarks in all possible ways.}
\item[(3)]{
Connect the remaining $\Ng-i$ gluons as in the gluons only case,
but keep cyclic and anti-cyclic tensors separately.
}
\item[(4)]{
Distribute the quark and gluon indices in all possible
ways among the different groupings.}
\end{list}
The number of color tensors  in this case grows slower than
$(\Ng+\Nq)!$ but faster than $(\Ng+\Nq)!/e$, again giving a factorial growth.

\section{Calculating the effect of gluon exchange}
\label{sec:Rules}

Below, the computational rules for gluon exchange will be derived, and
it will be seen that exchanging a gluon trivially gives an explicit linear 
combination of the basis tensors. There is thus no need to calculate scalar 
products of the resulting color structure after exchange, with the basis 
tensors. As calculating scalar products was the most cumbersome part in 
previous calculations, this represents a major improvement. 


\subsection{Computational rules}

In this section the computational rules for gluon exchange between the
basis tensors constructed in section \ref{sec:Construction} are derived.

Note that the quarks in closed quark loops are just products of the way of 
writing down the basis and not physical particles, a gluon is thus never
exchanged between the quarks in closed quark loops.

We also have to decide on a convention for the triple gluon vertex. The convention used is 
\begin{eqnarray}
\label{eq:gggSign}
& &f_{eig} \,\,\,\,\,\mbox{    with} \nonumber \\
& &e=\mbox{the external (incoming or outgoing) eikonal gluon index}\nonumber \\
& &i=\mbox{the internal (incoming or outgoing) eikonal gluon index} \\
& &g=\mbox{the soft exchange gluon index.}\nonumber
\end{eqnarray}
This convention has the advantage that the sign is independent of how
the diagram is drawn on a paper and whether a parton is incoming or
outgoing.

\subsubsection{Gluon exchange between two quarks or anti-quarks}

In the simplest case a gluon is exchanged between two external quarks
$q_1$ and $q_2$, which in general have $n$ and $m$
gluons attached respectively. Using \eqref{eq:tata}, the effect of gluon 
exchange between the quarks $q_1$ and $q_2$ in two different open quark
lines may be written

\begin{equation}
  \left(
    \begin{array}{ll}
      {{\qbar}_1 g_{11}... g_{1n} \q_1} &  \otimes \\
      {{\qbar}_2 g_{21}... g_{2m} \q_2} & 
    \end{array}
  \right)
\rightarrow
\frac{1}{2}
  \left(
    \begin{array}{ll}
      {{\qbar}_1 g_{11}... g_{1n} \q_2} & \otimes  \\
      {{\qbar}_2 g_{21}... g_{2m} \q_1} & 
    \end{array}
  \right)
-\frac{1}{2\Nc}
  \left(
    \begin{array}{ll}
      {{\qbar}_1 g_{11}... g_{1n} \q_1} &  \otimes \\
      {{\qbar}_2 g_{21}... g_{2m} \q_2} & 
    \end{array}
  \right).
  \label{eq:qq}
\end{equation}
where the notation
\begin{equation}
  \left(
    \begin{array}{ll}
      {{\qbar}_1 g_{11}... g_{1n} \q_1} &  \otimes \\
      {{\qbar}_2 g_{21}... g_{2m} \q_2} & 
    \end{array}
  \right)
  =t^{g_{11}}_{d_{11} \qbar_1}t^{g_{12}}_{d_{12}
    d_{11}}...t^{g_{1n}}_{q_1 d_{1n-1}}
  t^{g_{21}}_{d_{21} \qbar_2}t^{g_{22}}_{d_{22}
    d_{21}}...t^{g_{2m}}_{q_2 d_{2m-1}}
\end{equation}
is used.

If the gluon is instead exchanged between the external anti-quarks, the
indices on the quarks are kept whereas the indices on the anti-quarks
are exchanged. 

\subsubsection{Gluon exchange between quark and anti-quark}

Exchanging a gluon between a quark $q_1$ and an anti-quark
${\qbar}_2$ results in 

\begin{equation}
  \left(
    \begin{array}{ll}
      {{\qbar}_1 g_{11}... g_{1n} \q_1} &  \otimes \\
      {{\qbar}_2 g_{21}... g_{2m} \q_2} & 
    \end{array}
  \right)
\rightarrow
\frac{1}{2}
  \left(
    \begin{array}{ll}
      {{\qbar}_1 g_{11}... g_{1n}g_{21}...g_{2m}\q_2} & \otimes  \\
      {{\qbar}_2 \q_1} & 
    \end{array}
  \right)
-\frac{1}{2\Nc}
  \left(
    \begin{array}{ll}
      {{\qbar}_1 g_{11}... g_{1n} \q_1} & \otimes  \\
      {{\qbar}_2 g_{21}... g_{2m} \q_2} & 
    \end{array}
  \right).
  \label{eq:qqbar}
\end{equation}
The case where the involved quark and anti-quark are part of the same
quark line can be obtained by identifying  ${\qbar}_1$ and $q_2$ above.

\subsubsection{Gluon exchange between quark and  gluon}

To derive the effect of gluon exchange between a quark and a gluon we
use the relation \eqref{eq:fd} to rewrite the triple gluon
vertex. After this \eqref{eq:tata} is applied (and it is noted that the
$1/\Nc$ suppressed terms drop out). For gluon exchange between the quark
$q_1$ and the gluon $g_{2i}$ the result is:

\begin{equation}
  \left(
    \begin{array}{ll}
      {{\qbar}_1 g_{11}... g_{1n} \q_1 \;\; \otimes} &   \\
      {{\qbar}_2 g_{21}...g_{2i}... g_{2m} \q_2} & 
    \end{array}
  \right)
\rightarrow
-\frac{1}{2}
  \left(
    \begin{array}{ll}
      {{\qbar}_1 g_{11}... g_{1n}g_{2i+1}...g_{2m}\q_2} & \otimes  \\
      {{\qbar}_2 g_{21}...g_{2i}\q_1} & 
    \end{array}
  \right)
+\frac{1}{2}
  \left(
    \begin{array}{ll}
      {{\qbar}_1 g_{11}... g_{1n}g_{2i}...g_{2n} \q_2} & \otimes  \\
      {{\qbar}_2 g_{21}g_{2i-1} \q_1} & 
    \end{array}
  \right).
  \label{eq:qg}
\end{equation}
If, in the left hand side above, $g_{2i}$ is in a closed quark loop this
is accounted for by identifying $\qbar_2$ and $q_2$, and if the gluon
$g_{2i}$ is attached to the same quark-line as $\q_1$ this is taken
care of by identifying  $\qbar_1$ and $q_2$.

\subsubsection{Gluon exchange between anti-quark and gluon}

Employing the same calculational method as for $qg$ results in

\begin{equation}
  \left(
    \begin{array}{ll}
      {{\qbar}_1 g_{11}... g_{1n} \q_1 \;\;\otimes} &   \\
      {{\qbar}_2 g_{21}...g_{2i}... g_{2m} \q_2} & 
    \end{array}
  \right)
\rightarrow
\frac{1}{2}
  \left(
    \begin{array}{ll}
      {{\qbar}_1 g_{2i}...g_{2m}\q_2 \;\;\otimes} &   \\
      {{\qbar}_2 g_{21}...g_{2i-1}g_{11}...g_{2n}\q_1} & 
    \end{array}
  \right)
-\frac{1}{2}
  \left(
    \begin{array}{ll}
      {{\qbar}_1 g_{2i+1}... g_{2m}\q_2 \; \;\otimes} &   \\
      {{\qbar}_2 g_{21}...g_{2i}g_{11}...g_{1n}\q_1} & 
    \end{array}
  \right)
  \label{eq:qbarg}
\end{equation}
where again, if $g_{2i}$ initially is in a closed quark loop this is accounted
for by identifying $\qbar_2$ and $q_2$, and if the gluon $g_{2i}$ was
originally placed on the same quark line as $\qbar_1$ this is taken care of by
identifying  $\qbar_2$ and $q_1$.

\subsubsection{Gluon exchange between two external gluons}

To derive the effect on the basis vectors of exchanging a gluon between
two external gluons, two triple gluon vertices have to be replaced using
\eqref{eq:fd} 
and three gluon propagators have to be contracted using \eqref{eq:tata}.  
Again the
non-leading $\Nc$ terms drop out and the result of exchanging a gluon
between $g_{1i}$ and $g_{2j}$ is 

\begin{eqnarray}
 \label{eq:gg}
&  &\left(
    \begin{array}{ll}
      {{\qbar}_1 g_{11}...g_{1i}... g_{1n} \q_1} & \otimes  \\
      {{\qbar}_2 g_{21}...g_{2j}... g_{2m} \q_2} & 
    \end{array}
  \right)
\rightarrow \\
& &-\frac{1}{2}
  \left(
    \begin{array}{ll}
      {{\qbar}_1 g_{11}... g_{1i-1}g_{2j+1}...g_{2m}\q_2 \;\; \otimes} &  \\
      {{\qbar}_2 g_{21}... g_{2j-1}g_{2j}g_{1i}g_{1i+1}...g_{in}\q_1} & 
    \end{array}
  \right)
+\frac{1}{2}
  \left(
    \begin{array}{ll}
      {{\qbar}_1 g_{11}... g_{1i-1}g_{1i}g_{2j+1}...g_{2m}\q_2} & \otimes  \\
      {{\qbar}_2 g_{21}... g_{2j-1}g_{2j}g_{1i+1}...g_{1n}\q_1} & 
    \end{array}
  \right)\nonumber \\
& &+\frac{1}{2}
  \left(
    \begin{array}{ll}
      {{\qbar}_1 g_{11}... g_{1i-1}g_{2j}g_{2j+1}...g_{2m}\q_2} & \otimes  \\
      {{\qbar}_2 g_{21}... g_{2j-1}g_{1i}g_{1i+1}...g_{1n}\q_1} & 
    \end{array}
  \right)
-\frac{1}{2}
  \left(
    \begin{array}{ll}
      {{\qbar}_1 g_{11}... g_{1i-1}g_{1i}g_{2j}g_{2j+1}...g_{2m}\q_2} & \otimes  \\
      {{\qbar}_2 g_{21}... g_{2j-1}g_{1i+1}...g_{1n}\q_1} & 
    \end{array}
  \right). \nonumber
\end{eqnarray}
If one (or both) quark lines is (are) closed, then the corresponding
quarks are to be identified. If both gluons are part of the same quark
line, then identify $q_1\qbar_2$, and $q_2\qbar_1$ if the quark line is closed.

\section{Some explicit examples }
\label{sec:Example}

\subsection{$gg \rightarrow gg$}

As an explicit example of how the above strategy simplifies the problem
of keeping track of the color structure, the process of $g_1 g_2\rightarrow
g_3g_4$ will be considered in detail here. 
The soft anomalous dimension matrix for this case was first calculated in
\cite{Oderda:1999kr} and later, more elegantly in \cite{Dokshitzer:2005ig}.

\subsubsection{Construction of the basis}

To construct the basis the recipe outlined in section \ref{sec:GluonsOnly} is
followed, starting with finding all the ways of grouping the gluons:

\begin{list}{}{\setlength\leftmargin{10mm}}
\item[(1)]{
The four gluons can be grouped two and two $\{2,2\}$ or all four together
$\{4\}$.}

\item[(2)]{ When all four gluons are attached to the same quark line,
    $\{4\}$, the indices
can be placed in $(4-1)!=6$ different ways. However, due to the symmetry
between quarks and anti-quarks, clockwise and anti-clockwise gluon
rings only enter in one linear combination, giving three physical tensors:

\begin{eqnarray}
C^1_{g_1 g_2 g_3 g_4}&=&(g_1 g_2 g_3 g_4)+(g_4 g_3 g_2 g_1)
=\Tr[t^{g_1} t^{g_2} t^{g_3} t^{g_4}]+\Tr[t^{g_4} t^{g_3} t^{g_2} t^{g_1} ]\nonumber \\
C^{2}_{g_1 g_2 g_3 g_4}&=&(g_1 g_2 g_4 g_3)+(g_3 g_4 g_2 g_1)
=\Tr[t^{g_1} t^{g_2} t^{g_4} t^{g_3}]+\Tr[t^{g_3} t^{g_4}
t^{g_2} t^{g_1} ]\nonumber \\
C^{3}_{g_1 g_2 g_3 g_4}&=&(g_1  g_3 g_2 g_4)+(g_4 g_2 g_3 g_1)
=\Tr[t^{g_1} t^{g_3} t^{g_2} t^{g_4}]+\Tr[t^{g_4} t^{g_2}
t^{g_3} t^{g_1} ].\nonumber\\
& &
\end{eqnarray}
}
\item[(3a)]{ For the grouping $\{2,2\}$, the index order in the subgrouping doesn't
matter (since $\Tr[t^g_1 t^g_2]=\Tr[t^g_2 t^g_1]$). Each subgrouping thus only
gives rise to one physical tensor.}

\item[(3b,c)]{ The gluon indices $g_1,g_2,g_3,g_4$ may be split into the
subgroupings as $\{\{g_1,g_2\},\{g_3,g_4\}\}$, $\{\{g_1,g_3\},\{g_2,g_4\}\}$ and
$\{\{g_1,g_4\},\{g_2,g_3\}\}$, 
giving three basis tensors

\begin{eqnarray}
  C^4_{g_1 g_2 g_3 g_4}&=&(g_1 g_2)(g_3 g_4)
  =\Tr[t^{g_1} t^{g_2}]\Tr[t^{g_3} t^{g_4}]
  =\left( \frac{1}{2} \right)^2 \delta_{g_1g_2}\delta_{g_3g_4}\nonumber \\
  C^5_{g_1 g_2 g_3 g_4}&=&(g_1 g_3)(g_2 g_4)
  =\Tr[t^{g_1} t^{g_3}]\Tr[t^{g_2} t^{g_4}]=
  \left( \frac{1}{2} \right)^2 \delta_{g_1 g_3}\delta_{g_2 g_4}\nonumber \\
  C^{6}_{g_1 g_2 g_3 g_4}&=&(g_1 g_4)(g_2 g_3)
  =\Tr[t^{g_1} t^{g_4}]\Tr[t^{g_2} t^{g_3}]=
  \left( \frac{1}{2} \right)^2 \delta_{g_1g_4}\delta_{g_2g_3}.
\end{eqnarray}
}
\end{list}

\subsubsection{Calculation of soft anomalous dimension matrix}
\label{sec:Gamgggg}

As previously noted, once the effect of gluon exchange between $g_1$
and $g_2$ is calculated the effect of gluon exchange between any other
gluons may be deduced. There may thus at most be $\Nbasis$ different
situations to keep track of. However, this number will in general be
further reduced due to the irrelevance of non-participating indices. For
example the effect of gluon exchange between $g_1$ and $g_2$ on
$C^1_{g_1g_2g_3g_4}$ is the same as the effect on $C^2_{g_1g_2g_3g_4}$.
Thus, the physically different situations are:

\begin{list}{}{\setlength\leftmargin{10mm}}

\item[(1)]{A gluon is exchanged between two neighboring gluons on a quark
    ring with four gluons attached (for example gluon 1 and 2 on $C^1$).

Applying \eqref{eq:gg} to the first half of $C^{1}_{g_1g_2g_3g_4}$
with the identification $g_1 \rightarrow g_{12}=g_{1i}$, $g_2\rightarrow
g_{21}=g_{2j}$, $g_3\rightarrow g_{22}$, $g_4\rightarrow g_{11}$ and
$\qbar_1=q_2$, $\qbar_2=q_1$
gives 

\begin{eqnarray}
  (g_1g_2g_3g_4) 
  &\rightarrow& -\frac{1}{2}
  (g_1g_2)(g_3g_4)-\frac{\Nc}{2}(g_1g_2g_3g_4).
\end{eqnarray}
Similarly application to the second half results in 
\begin{eqnarray}
  (g_1g_4g_3g_2) 
  &\rightarrow& -\frac{1}{2} (g_1g_2)(g_3g_4)-\frac{\Nc}{2}(g_1g_4g_3g_2),
\end{eqnarray}
and it may be concluded that 
\begin{eqnarray}
  C^{1}_{g_1g_2g_3g_4}
  &\rightarrow& -\frac{2}{2} C^{4}_{g_1g_2g_3g_4} -\frac{\Nc}{2}C^{1}_{g_1g_2g_3g_4}.
  \label{eq:Close}
\end{eqnarray}
}

\item[(2)]{A gluon can be exchanged between two next to neighboring
    gluons. 
    In this case we get for an exchange between  $g_1$ and $g_2$
on $C^{3}_{g_1g_2g_3g_4}$
\begin{eqnarray}
  C^{3}_{g_1g_2g_3g_4}
  &\rightarrow&  C^{5}_{g_1g_2g_3g_4} +C^{6}_{g_1g_2g_3g_4}.
\end{eqnarray}
}

\item[(3)]{A gluon may be exchanged between the gluons attached to a two
gluon ring, such as $g_1$ and $g_2$ on $C^{4}_{g_1g_2g_3g_4}$. 
This just gives a factor $\Nc$ multiplying the old tensor, for example for
gluon exchange between $g_1$ and $g_2$ on $C^{4}_{g_1g_2g_3g_4}$
\begin{eqnarray}
  C^{4}_{g_1g_2g_3g_4}
  &\rightarrow& \Nc C^{4}_{g_1g_2g_3g_4}.
\end{eqnarray}
 }

\item[(4)]{A gluon may be exchange between two gluons attached to different two
gluon rings such as $g_1$ and $g_2$ in  $C^{6}_{g_1g_2g_3g_4}$,
giving 

\begin{eqnarray}
  C^{6}_{g_1g_2g_3g_4}
  &\rightarrow& -\frac{1}{2} C^{1}_{g_1g_2g_3g_4}
  +\frac{1}{2}C^{3}_{g_1g_2g_3g_4}.
  \label{eq:DiffRings}
\end{eqnarray}
}
\end{list}

The above information may be combined into a matrix describing the color
algebra part for gluon exchange between $g_1$ and $g_2$
\begin{eqnarray}
  \mathbf{C}^{12}_{gg \to gg}=
  \left(
  \begin{array}{llllll}
    \frac{-\Nc}{2} & 0 & 0 & 0 & 0 &- \frac{1}{2}  \\
    0 & \frac{-\Nc}{2} & 0 & 0 & -\frac{1}{2} &  0\\
    0 & 0 & 0 & 0 & \frac{1}{2} & \frac{1}{2} \\
    -1 & -1 & 0 & \Nc & 0 & 0 \\
    0 & 0 & 1  & 0 & 0 & 0 \\
    0 & 0 & 1 & 0 & 0 & 0
  \end{array}
  \right).\nonumber
  \label{eq:Ggggg12}
\end{eqnarray}

As the gluons $g_3$ and $g_4$ have the same relationship to each other in
the basis as $g_1$ and $g_2$, the color structure of the soft anomalous 
dimension matrix will be the same 
$\mathbf{C}^{34}_{gg \to gg}=\mathbf{C}^{12}_{gg \to gg}$. Similarly
 $\mathbf{C}^{14}_{gg \to gg}=\mathbf{C}^{23}_{gg \to gg}$ and 
$\mathbf{C}^{24}_{gg \to gg}=\mathbf{C}^{13}_{gg\to gg}$.

The contributions  $\mathbf{C}^{14}_{gg \to gg}$ and $\mathbf{C}^{24}_{gg \to gg}$ may be 
calculated by using the results in Eqs.
(\ref{eq:Close}-\ref{eq:DiffRings}) and relabeling indices.
Letting 
$T=\Omega_{12}+\Omega_{34}$, $U=\Omega_{13}+\Omega_{24}$ and 
$V=\Omega_{14}+\Omega_{23}$ 
be the phase space integrals the result can be written

\begin{eqnarray}
\label{eq:gggg}
& &\G_{gg \to gg}=\\
& &\left(
\begin{array}{cccccc}
 -\frac{1}{2} \Nc (T+V) & 0 & 0 & \frac{U-V}{2} & 0 & \frac{U-T}{2} \\
 0 & -\frac{1}{2} \Nc (T+U) & 0 & \frac{V-U}{2} & \frac{V-T}{2} & 0 \\
 0 & 0 & -\frac{1}{2} \Nc (U+V) & 0 & \frac{T-V}{2} & \frac{T-U}{2} \\
 U-T & V-T & 0 & -\Nc T & 0 & 0 \\
 0 & V-U & T-U & 0 & -\Nc U & 0 \\
 U-V & 0 & T-V & 0 & 0 & -\Nc V
\end{array}
\right).\nonumber 
\end{eqnarray}

To obtain physical results the scalar product matrix
\newpage
\begin{eqnarray}
\label{eq:SPgggg}
& &{\Sv}_{gg \to gg}=\\
& &
\left(
\begin{array}{cccccc}
 \frac{\Nc^6-3 \Nc^4+8 \Nc^2-6}{8 \Nc^2} & \frac{-\Nc^4+4 \Nc^2-3}{4 \Nc^2} &
   \frac{-\Nc^4+4 \Nc^2-3}{4 \Nc^2} & \frac{\left(\Nc^2-1\right)^2}{8 \Nc} &
   \frac{1-\Nc^2}{8 \Nc} & \frac{\left(\Nc^2-1\right)^2}{8 \Nc} \\
 \frac{-\Nc^4+4 \Nc^2-3}{4 \Nc^2} & \frac{\Nc^6-3 \Nc^4+8 \Nc^2-6}{8 \Nc^2} &
   \frac{-\Nc^4+4 \Nc^2-3}{4 \Nc^2} & \frac{\left(\Nc^2-1\right)^2}{8 \Nc} &
   \frac{\left(\Nc^2-1\right)^2}{8 \Nc} & \frac{1-\Nc^2}{8 \Nc} \\
 \frac{-\Nc^4+4 \Nc^2-3}{4 \Nc^2} & \frac{-\Nc^4+4 \Nc^2-3}{4 \Nc^2} & \frac{\Nc^6-3
   \Nc^4+8 \Nc^2-6}{8 \Nc^2} & \frac{1-\Nc^2}{8 \Nc} &
   \frac{\left(\Nc^2-1\right)^2}{8 \Nc} & \frac{\left(\Nc^2-1\right)^2}{8 \Nc}
   \\
 \frac{\left(\Nc^2-1\right)^2}{8 \Nc} & \frac{\left(\Nc^2-1\right)^2}{8 \Nc} &
   \frac{1-\Nc^2}{8 \Nc} & \frac{\left(\Nc^2-1\right)^2}{16} & \frac{
   \left(\Nc^2-1\right)}{16} & \frac{\left(\Nc^2-1\right)}{16} \\
 \frac{1-\Nc^2}{8 \Nc} & \frac{\left(\Nc^2-1\right)^2}{8 \Nc} &
   \frac{\left(\Nc^2-1\right)^2}{8 \Nc} & \frac{\left(\Nc^2-1\right)}{16} &
   \frac{\left(\Nc^2-1\right)^2}{16} & \frac{\left(\Nc^2-1\right)}{16}
   \\
 \frac{\left(\Nc^2-1\right)^2}{8 \Nc} & \frac{1-\Nc^2}{8 \Nc} &
   \frac{\left(\Nc^2-1\right)^2}{8 \Nc} & \frac{\left(\Nc^2-1\right)}{16} &
   \frac{\left(\Nc^2-1\right)}{16} & \frac{\left(\Nc^2-1\right)^2}{16}
\end{array}
\right),\nonumber
\end{eqnarray}
calculated using \eqref{eq:SP},
is also needed. This matrix contains ${\Nbasis}^2$ entries, 
however, closer consideration reveals that only six of them
correspond to different contractions.

It is worth remarking on the leading $\Nc$ behavior of
\eqref{eq:gggg}. The computational rules in Eqs. (\ref{eq:qq}-\ref{eq:gg})
contain no positive power of $\Nc$. Thus the $\Nc$ in
\eqref{eq:gggg} must come from closed quark loops. The only way to get a
closed quark loop is to exchange a gluon between two neighboring partons
attached to the same quark-line, i.e. only "color neighbors"
radiate in the $\Nc \rightarrow \infty$ limit. 
The result after exchange contains a factor $\Nc$ multiplying the old
color structure. Leading $\Nc$ contributions will therefore always be
diagonal in the present basis.

This is in close resemblance with the Dipole Cascade Model and
the original Ariadne program in which only neighboring pairs of partons, 
dipoles, radiate  
\cite{Gustafson:1986db,Gustafson:1987rq,Pettersson:1988zu,Lonnblad:1992tz}.
In particular for gluon radiation from $e^+
e^- \rightarrow q \qbar$ the leading $\Nc$ piece should come from
neighboring partons. It is cautioned, however, that there are many more
non-leading $\Nc$ contributions, than leading pieces, as there (in
general) are many more non-neighboring partons. 

It is also worth remarking that although there are scalar products
between different basis tensors that are suppressed by only one power of
$\Nc$, these scalar products are never between different fully connected
topologies, i.e. never between tree level QCD gluon amplitudes.

\subsection{$q \qbar \rightarrow q \qbar g$}

As an example of a process containing both quarks and gluons we consider the 
color structure needed for gluon resummation for
$q_1 \qbar_2 \rightarrow q_3 \qbar_4 g_5$. This color structure is
important for (among other things) QCD corrections to the production of
$W$'s decaying leptonically and being accompanied by three jets
\cite{Berger:2009zg}.

The result of constructing color tensors as outlined in section \ref{sec:Both}
is

\begin{eqnarray}
  C^1_{q_1 q_2 q_3 q_4 g_5}&=&(\qbar_4 g_5 q_3)(\qbar_1 q_2)
  =t^{g_5}_{q_3 q_4} \delta_{q_1 q_2} \nonumber \\
  C^2_{q_1 q_2 q_3 q_4 g_5}&=&(\qbar_4 q_3)(\qbar_1 g_5 q_2)
  =\delta_{q_3 q_4} t^{g_5}_{q_2 q_1} \nonumber \\
  C^3_{q_1 q_2 q_3 q_4 g_5}&=&(\qbar_4 g_5 q_2)(\qbar_1 q_3)
  =t^{g_5}_{q_2 q_4} \delta_{q_3 q_1} \nonumber \\
  C^4_{q_1 q_2 q_3 q_4 g_5}&=&(\qbar_2 q_4)(\qbar_1 g_5 q_3)
  =\delta_{q_2 q_4} t^{g_5}_{q_3 q_1}.
\end{eqnarray}
Exchanging gluons between the partons in all possible ways results in a 
leading color diagonal part, a $1/N_c$ suppressed off-diagonal part and a 
$1/\Nc^2$ suppressed diagonal part:
{\small
\begin{eqnarray}
  & &\G_{q \qbar \rightarrow q \qbar g} =\nonumber\\
  & &\frac{N_c}{2} \mbox{Diagonal}[
  \Omega _{12}+\Omega _{35}-\Omega_{45},\;
  -\Omega_{15}+\Omega _{25}+\Omega _{34},\;
  \Omega _{13}+\Omega_{25}-\Omega _{45},\;
  -\Omega _{15}+\Omega _{24}+\Omega_{35}]\nonumber \\
  & &+  \frac{1}{2}\left(
    \begin{array}{cccc}
      0 & 0 & \Omega _{12}+\Omega _{15}+\Omega _{23}+\Omega _{35} \;\;&
      \Omega _{12}+\Omega _{14}-\Omega _{25}-\Omega _{45} \\
      0 & 0 & \Omega _{14}-\Omega _{15}+\Omega _{34}-\Omega _{35} \;\; &
      \Omega _{23}+\Omega _{25}+\Omega _{34}+\Omega _{45} \\
      \Omega _{13}+\Omega _{15}+\Omega _{23}+\Omega _{25} \;\;& \Omega
      _{13}+\Omega _{14}-\Omega _{35}-\Omega _{45} & 0 & 0 \\
      \Omega _{14}-\Omega _{15}+\Omega _{24}-\Omega _{25} \;\;& \Omega
      _{23}+\Omega _{24}+\Omega _{35}+\Omega _{45} & 0 & 0
    \end{array}
  \right) \nonumber \\
   & &-\frac{1} {2 N_c} 
   (\Omega _{12}+\Omega _{13}+\Omega _{14}+\Omega
   _{23}+\Omega _{24}+\Omega _{34})\;
\mbox{Diagonal}[1,\;1,\;1,\;1].\nonumber
\end{eqnarray}}
\vspace*{-15mm}
\begin{equation}
\end{equation}
Again, as we are working in a non-orthogonal basis, all scalar products are 
needed
\begin{equation}
\Sv_{q \qbar \rightarrow q \qbar g}=
\left(
\begin{array}{cccc}
 \frac{1}{2} \Nc \left(\Nc^2-1\right) & 0 & \frac{1}{2} \left(\Nc^2-1\right)
   & \frac{1}{2} \left(\Nc^2-1\right) \\
 0 & \frac{1}{2} \Nc \left(\Nc^2-1\right) & \frac{1}{2} \left(\Nc^2-1\right)
   & \frac{1}{2} \left(\Nc^2-1\right) \\
 \frac{1}{2} \left(\Nc^2-1\right) & \frac{1}{2} \left(\Nc^2-1\right) &
   \frac{1}{2} \Nc \left(\Nc^2-1\right) & 0 \\
 \frac{1}{2} \left(\Nc^2-1\right) & \frac{1}{2} \left(\Nc^2-1\right) & 0 &
   \frac{1}{2} \Nc \left(\Nc^2-1\right)
\end{array}
\right).
\end{equation}

\subsection{$q \qbar \rightarrow ggg$}

The other color structure relevant for $W$ plus three jets is that of 
$q \qbar \rightarrow g g g$. In this case an 11-dimensional matrix is
needed to describe the color space (reducing to 10 for $\Nc=3$). 
These results have been calculated and are electronically attached to this 
submission. Again there is a diagonal leading $\Nc$ part, an off-diagonal
part with relative suppression $1/\Nc$ and a $1/\Nc^2$ suppressed diagonal contribution.

\section{Conclusions}
\label{sec:Conclusions}

In the present paper a general recipe for constructing bases capable
of dealing with the color structure needed for resummation for any
number of colored partons has been presented. This in itself is a step
forward. In addition the suggested bases are argued to have relatively nice
computational properties.
The bases are obtained from the $\Nc=\infty$ case by splitting gluons in 
$q \qbar$ pairs and 
connecting color lines in all possible ways. The bases thus constructed
will therefore neither be normalized or orthogonal for $\Nc=3$, 
but they will span the space and have the property that gluon exchanges
between any pair of external partons directly, i.e. without taking scalar
products, result in linear combinations of basis vectors. Furthermore,
as can be seen from the computational rules in
Eqs. (\ref{eq:qq}-\ref{eq:gg}),
the result after gluon exchange contains at most four (often two or one) basis
vectors, giving relatively sparse soft anomalous dimension matrices.

The fact that there is no need to calculate scalar products to decompose
the tensors resulting after gluon exchange is a major advantage. 
Otherwise there would, for 
each of the $\Nparton (\Nparton-1)/2$ possible gluon exchanges, be
$\sim \Nbasis^2 \sim (\Nparton!)^2$ (cf. section 
\ref{sec:QuarksOnly}-\ref{sec:Both}) scalar products to calculate.
In addition, the computational time for calculating the effect of gluon
exchange is further reduced, as the
constructed bases maximally exploit the symmetry w.r.t. external parton
indices. \textit{All} indices corresponding to the same kind of parton
enter the basis on equal footing. Therefore, for example, once the
effect of gluon exchange between any pair of gluons has been calculated,
the effect of gluon exchange between any other can be obtained by
relabeling of indices, corresponding to a renumbering of tensors. 
One thus at most has to calculate six
($gg$, $qq$, $\qbar \qbar$, or $q \qbar$ to $qg$ and $\qbar g$) 
different contributions to $\G$. For a hard scattering amplitude with only
gluons it is enough to calculate one.
 

From this it is clear that the major computational effort lies in
computing the $\sim \Nbasis^2$ scalar products between all the basis
vectors. 
So far, it thus looks as if the calculational effort is reduced by a
factor $\sim \Nparton (\Nparton-1)/2$ (times a factor coming from the fact
that the basis vectors are simpler to take scalar products of).
However, also in the case of calculating scalar products between basis
tensors, the equal footing of the indices
comes to rescue. This is so, as only the topology of the contraction,
and not the labeling of indices, is important for determining the
scalar product. 


Apart from the nicer scaling properties, the suggested bases have the
advantage of being orthogonal and giving rise to diagonal soft
anomalous dimension matrices in the $\Nc \rightarrow \infty$
limit. This enables a more straight forward comparison to event
generators, tending to keep only the leading $\Nc$ contribution.

\section*{Acknowledgments}
I am thankful to Stefan Gieseke, Gösta Gustafson and Mike Seymour for inspiring
discussions.
This work was supported by the EU, though a Marie Curie Experienced
Researcher fellowship of the MCnet Research Training network, contract MRTN-CT-2006-035606.

\bibliographystyle{utcaps}  
\bibliography{RRefs}

\providecommand{\href}[2]{#2}\begingroup\raggedright\begin{thebibliography}{10%
}\itemsep 0mm

\bibitem{Lipatov:1974qm}
L.~N. Lipatov {\em Sov. J. Nucl. Phys.} {\bf 20} (1975)
94--102.

\bibitem{Gribov:1972ri}
V.~N. Gribov and L.~N. Lipatov {\em Sov. J. Nucl. Phys.} {\bf 15} (1972)
438--450.

\bibitem{Altarelli:1977zs}
G.~Altarelli and G.~Parisi {\em Nucl. Phys.} {\bf B126} (1977)
298.

\bibitem{Dokshitzer:1977sg}
Y.~L. Dokshitzer {\em Sov. Phys. JETP} {\bf 46} (1977)
641--653.

\bibitem{Kidonakis:1998nf}
N.~Kidonakis, G.~Oderda, and G.~Sterman {\em Nucl. Phys.} {\bf B531} (1998)
  365--402,
\href{http://www.arXiv.org/abs/hep-ph/9803241}{{\tt hep-ph/9803241}}.

\bibitem{Sjodahl:2008fz}
M.~Sjodahl {\em JHEP} {\bf 12} (2008) 083,
\href{http://www.arXiv.org/abs/0807.0555}{{\tt 0807.0555}}.

\bibitem{Seymour:2005ze}
M.~H. Seymour {\em JHEP} {\bf 10} (2005) 029,
\href{http://www.arXiv.org/abs/hep-ph/0508305}{{\tt hep-ph/0508305}}.

\bibitem{Seymour:2008xr}
M.~H. Seymour and M.~Sjodahl {\em JHEP} {\bf 12} (2008) 066,
\href{http://www.arXiv.org/abs/0810.5756}{{\tt 0810.5756}}.

\bibitem{Botts:1989kf}
J.~Botts and G.~Sterman {\em Nucl. Phys.} {\bf B325} (1989)
62.

\bibitem{Sotiropoulos:1993rd}
M.~G. Sotiropoulos and G.~Sterman {\em Nucl. Phys.} {\bf B419} (1994) 59--76,
\href{http://www.arXiv.org/abs/hep-ph/9310279}{{\tt hep-ph/9310279}}.

\bibitem{Contopanagos:1996nh}
H.~Contopanagos, E.~Laenen, and G.~Sterman {\em Nucl. Phys.} {\bf B484} (1997)
  303--330,
\href{http://www.arXiv.org/abs/hep-ph/9604313}{{\tt hep-ph/9604313}}.

\bibitem{Oderda:1999kr}
G.~Oderda {\em Phys. Rev.} {\bf D61} (2000) 014004,
\href{http://www.arXiv.org/abs/hep-ph/9903240}{{\tt hep-ph/9903240}}.

\bibitem{Appleby:2003hp}
R.~B. Appleby
\href{http://www.arXiv.org/abs/hep-ph/0311210}{{\tt hep-ph/0311210}}.

\bibitem{Dokshitzer:2005ig}
Y.~L. Dokshitzer and G.~Marchesini {\em JHEP} {\bf 01} (2006) 007,
\href{http://www.arXiv.org/abs/hep-ph/0509078}{{\tt hep-ph/0509078}}.

\bibitem{Kyrieleis:2005dt}
A.~Kyrieleis and M.~H. Seymour {\em JHEP} {\bf 01} (2006) 085,
\href{http://www.arXiv.org/abs/hep-ph/0510089}{{\tt hep-ph/0510089}}.

\bibitem{Oderda:1998en}
G.~Oderda and G.~Sterman {\em Phys. Rev. Lett.} {\bf 81} (1998) 3591--3594,
\href{http://www.arXiv.org/abs/hep-ph/9806530}{{\tt hep-ph/9806530}}.

\bibitem{Berger:2001ns}
C.~F. Berger, T.~Kucs, and G.~Sterman {\em Phys. Rev.} {\bf D65} (2002) 094031,
\href{http://www.arXiv.org/abs/hep-ph/0110004}{{\tt hep-ph/0110004}}.

\bibitem{Berger:2002ig}
C.~F. Berger, T.~Kucs, and G.~Sterman {\em Int. J. Mod. Phys.} {\bf A18} (2003)
  4159--4168,
\href{http://www.arXiv.org/abs/hep-ph/0212343}{{\tt hep-ph/0212343}}.

\bibitem{Appleby:2002ke}
R.~B. Appleby and M.~H. Seymour {\em JHEP} {\bf 12} (2002) 063,
\href{http://www.arXiv.org/abs/hep-ph/0211426}{{\tt hep-ph/0211426}}.

\bibitem{Derrick:1995pb}
{\bf ZEUS} Collaboration, M.~Derrick {\em et al.} {\em Phys. Lett.} {\bf B369}
  (1996) 55--68,
\href{http://www.arXiv.org/abs/hep-ex/9510012}{{\tt hep-ex/9510012}}.

\bibitem{Abe:1997ie}
{\bf CDF} Collaboration, F.~Abe {\em et al.} {\em Phys. Rev. Lett.} {\bf 80}
  (1998)
1156--1161.

\bibitem{Abe:1998ip}
{\bf CDF} Collaboration, F.~Abe {\em et al.} {\em Phys. Rev. Lett.} {\bf 81}
  (1998)
5278--5283.

\bibitem{Abbott:1998jb}
{\bf D0} Collaboration, B.~Abbott {\em et al.} {\em Phys. Lett.} {\bf B440}
  (1998) 189--202,
\href{http://www.arXiv.org/abs/hep-ex/9809016}{{\tt hep-ex/9809016}}.

\bibitem{Adloff:2002em}
{\bf H1} Collaboration, C.~Adloff {\em et al.} {\em Eur. Phys. J.} {\bf C24}
  (2002) 517--527,
\href{http://www.arXiv.org/abs/hep-ex/0203011}{{\tt hep-ex/0203011}}.

\bibitem{Forshaw:2008cq}
J.~R. Forshaw, A.~Kyrieleis, and M.~H. Seymour {\em JHEP} {\bf 09} (2008) 128,
\href{http://www.arXiv.org/abs/0808.1269}{{\tt 0808.1269}}.

\bibitem{Gardi:2009qi}
E.~Gardi and L.~Magnea {\em JHEP} {\bf 03} (2009) 079,
\href{http://www.arXiv.org/abs/0901.1091}{{\tt 0901.1091}}.

\bibitem{Becher:2009cu}
T.~Becher and M.~Neubert {\em Phys. Rev. Lett.} {\bf 102} (2009) 162001,
\href{http://www.arXiv.org/abs/0901.0722}{{\tt 0901.0722}}.

\bibitem{Becher:2009qa}
T.~Becher and M.~Neubert {\em JHEP} {\bf 06} (2009) 081,
\href{http://www.arXiv.org/abs/0903.1126}{{\tt 0903.1126}}.

\bibitem{Becher:2009kw}
T.~Becher and M.~Neubert {\em Phys. Rev.} {\bf D79} (2009) 125004,
\href{http://www.arXiv.org/abs/0904.1021}{{\tt 0904.1021}}.

\bibitem{Mitov:2009sv}
A.~Mitov, G.~Sterman, and I.~Sung {\em Phys. Rev.} {\bf D79} (2009) 094015,
\href{http://www.arXiv.org/abs/0903.3241}{{\tt 0903.3241}}.

\bibitem{Sjostrand:2006za}
T.~Sjostrand, S.~Mrenna, and P.~Skands {\em JHEP} {\bf 05} (2006) 026,
\href{http://www.arXiv.org/abs/hep-ph/0603175}{{\tt hep-ph/0603175}}.

\bibitem{Sjostrand:2007gs}
T.~Sjostrand, S.~Mrenna, and P.~Skands {\em Comput. Phys. Commun.} {\bf 178}
  (2008) 852--867,
\href{http://www.arXiv.org/abs/0710.3820}{{\tt 0710.3820}}.

\bibitem{Corcella:2002jc}
G.~Corcella {\em et al.}
\href{http://www.arXiv.org/abs/hep-ph/0210213}{{\tt hep-ph/0210213}}.

\bibitem{Bahr:2008pv}
M.~Bahr {\em et al.} {\em Eur. Phys. J.} {\bf C58} (2008) 639--707,
\href{http://www.arXiv.org/abs/0803.0883}{{\tt 0803.0883}}.

\bibitem{Gleisberg:2003xi}
T.~Gleisberg {\em et al.} {\em JHEP} {\bf 02} (2004) 056,
\href{http://www.arXiv.org/abs/hep-ph/0311263}{{\tt hep-ph/0311263}}.

\bibitem{'tHooft:1973jz}
G.~'t~Hooft {\em Nucl. Phys.} {\bf B72} (1974)
461.

\bibitem{Gustafson:1982ws}
G.~Gustafson {\em Z. Phys.} {\bf C15} (1982)
155--160.

\bibitem{Dasgupta:2001sh}
M.~Dasgupta and G.~P. Salam {\em Phys. Lett.} {\bf B512} (2001) 323--330,
\href{http://www.arXiv.org/abs/hep-ph/0104277}{{\tt hep-ph/0104277}}.

\bibitem{Dasgupta:2002bw}
M.~Dasgupta and G.~P. Salam {\em JHEP} {\bf 03} (2002) 017,
\href{http://www.arXiv.org/abs/hep-ph/0203009}{{\tt hep-ph/0203009}}.

\bibitem{Forshaw:2006fk}
J.~R. Forshaw, A.~Kyrieleis, and M.~H. Seymour {\em JHEP} {\bf 08} (2006) 059,
\href{http://www.arXiv.org/abs/hep-ph/0604094}{{\tt hep-ph/0604094}}.

\bibitem{Forshaw:2009sf}
J.~R. Forshaw and M.~H. Seymour
\href{http://www.arXiv.org/abs/0901.3037}{{\tt 0901.3037}}.

\bibitem{MertAybat:2006mz}
S.~Mert~Aybat, L.~J. Dixon, and G.~Sterman {\em Phys. Rev.} {\bf D74} (2006)
  074004,
\href{http://www.arXiv.org/abs/hep-ph/0607309}{{\tt hep-ph/0607309}}.

\bibitem{Paton:1969je}
J.~E. Paton and H.-M. Chan {\em Nucl. Phys.} {\bf B10} (1969)
516--520.

\bibitem{Mangano:1987xk}
M.~L. Mangano, S.~J. Parke, and Z.~Xu {\em Nucl. Phys.} {\bf B298} (1988)
653.

\bibitem{Berends:1987cv}
F.~A. Berends and W.~Giele {\em Nucl. Phys.} {\bf B294} (1987)
700.

\bibitem{Nagy:2007ty}
Z.~Nagy and D.~E. Soper {\em JHEP} {\bf 09} (2007) 114,
\href{http://www.arXiv.org/abs/0706.0017}{{\tt 0706.0017}}.

\bibitem{Gustafson:1986db}
G.~Gustafson {\em Phys. Lett.} {\bf B175} (1986)
453.

\bibitem{Gustafson:1987rq}
G.~Gustafson and U.~Pettersson {\em Nucl. Phys.} {\bf B306} (1988)
746.

\bibitem{Pettersson:1988zu}
U.~Pettersson. LU-TP-88-5.

\bibitem{Lonnblad:1992tz}
L.~Lonnblad {\em Comput. Phys. Commun.} {\bf 71} (1992)
15--31.

\bibitem{Berger:2009zg}
C.~F. Berger {\em et al.}
\href{http://www.arXiv.org/abs/0902.2760}{{\tt 0902.2760}}.

\end{thebibliography}\endgroup

\end{document}